\documentclass[12pt]{article}
\usepackage{epsfig}
\usepackage{amsmath}
\usepackage{amsfonts}
\usepackage{amssymb}

\begin{document}
\pagestyle{empty}
\begin{flushright}
UMN-TH-2712/08
\end{flushright}
\vspace*{5mm}

\begin{center}
{\LARGE \bf  The Soft-Wall Standard Model}

\vspace{1.0cm}

{\sc Brian Batell$^{a,}$}\footnote{E-mail:  batell@physics.umn.edu}$^{,\dagger}$
{\sc Tony Gherghetta$^{b,}$}\footnote{E-mail:  tgher@unimelb.edu.au}
{\small and}
{\sc Daniel Sword$^{a,}$}\footnote{E-mail:  sword@physics.umn.edu\\
$^{\hspace{.5cm}\dagger}$Address after 1 September 2008: Perimeter Institute for Theoretical Physics, Waterloo, Ontario N2L 2Y5, Canada}
\\
\vspace{.5cm}
{\it\small {$^a$School of Physics and Astronomy, University of Minnesota,\\
Minneapolis, MN 55455, USA}}\\
{\it\small {$^b$School of Physics, University of Melbourne, Victoria 3010,
Australia}}\\
\end{center}

\vspace{1cm}
\begin{abstract}
We explore the possibility of modeling electroweak physics in a warped extra dimension with a soft wall. 
The infrared boundary is replaced with a smoothly varying dilaton field that provides a dynamical spacetime 
cutoff. We analyze gravity, gauge fields, and fermions in the soft-wall background and obtain a discrete 
spectrum of Kaluza-Klein states which can exhibit linear Regge-like behavior. Bulk Yukawa interactions give
rise to nonconstant fermion mass terms, leading to fermion localization in the soft-wall background and a possible explanation of the Standard Model flavor structure. Furthermore we construct electroweak models with custodial symmetry, where the gauge symmetry is broken with a bulk Higgs condensate. The electroweak constraints are not as stringent as in hard-wall models, allowing Kaluza-Klein masses of order the TeV scale.
\end{abstract}

\vfill
\begin{flushleft}
\end{flushleft}
\eject
\pagestyle{empty}
\setcounter{page}{1}
\setcounter{footnote}{0}
\pagestyle{plain}

\section{Introduction}

The warped extra dimension framework provides a compelling geometrical understanding of a number of mysteries 
left unexplained by the Standard Model (SM), most notably the hierarchy problem and the flavor puzzle. 
In the original Randall-Sundrum model~\cite{rs}, a slice of AdS bounded by ultraviolet (UV) and infrared 
(IR) branes was used to solve the hierarchy problem. This setup was subsequently generalized by placing SM fields in the bulk \cite{gb1,gb2,ferm,chang,gp} in order to address flavor issues.
Fermion mass hierarchies result from the wavefunction overlap of SM fermions with an IR localized Higgs \cite{ferm, gp, huber}. This leads to a fermion geography in the fifth dimension which also naturally suppresses the scale of generic higher-dimension operators that mediate dangerous flavor-changing processes \cite{gp, huber}.
Furthermore, there exists a built-in ``GIM'' mechanism suppressing induced flavor changing neutral currents (FCNC), a result of the near-universality of the couplings between the SM fermions and excited Kaluza-Klein (KK) gauge modes~\cite{gp,gim}. In addition, by the AdS/CFT correspondence~\cite{adscft,holo}, these extra-dimensional models provide a weakly-coupled holographic description of nontrivial strong gauge dynamics responsible for electroweak symmetry breaking and flavor physics (for a review, see \cite{review}).

A basic feature of warped phenomenological models is the existence of an IR brane at which the warped dimension abruptly ends.  This breaks the conformal symmetry, generating four-dimensional (4D) particle states with a KK/composite mass spectrum $m_n^2\sim n^2$. But this hard-wall representation of the IR brane represents just one way to break conformal symmetry. Instead a more general approach is to replace the IR brane with a so-called soft wall, in which conformal symmetry is smoothly broken by a dilaton field, providing a dynamical cutoff to spacetime along the fifth dimension. This allows for a variety of KK mass spectra to be generated and, since there is no longer an IR brane, forces all IR brane fields to be five-dimensional (5D) bulk fields. Thus, from the dual holographic description, any operator of finite dimension responsible for conformal (or other) symmetry breaking can be modeled in the soft-wall background.

The soft-wall warped dimension was first proposed to model the Regge behavior of highly excited mesons in 
AdS/QCD models~\cite{adsqcd}. In analogy with this QCD application, the soft wall can be used to model the possible underlying dynamics of electroweak physics. Since this dynamics is unknown, a much larger set of possibilities can be studied for electroweak physics, leading to a variety of qualitatively distinct phenomenology. 
In particular, an application of the soft-wall warped dimension to electroweak breaking in gauge-Higgs unification models leads to less severe constraints from electroweak precision tests, allowing for KK gauge modes generically as light as 2 TeV or less~\cite{fpv}.

In this work we will study the Standard Model in a soft-wall warped dimension as a way to model the possible
underlying dynamics of electroweak physics.
A concrete 5D gravity model, similar to the dynamical AdS/QCD model of Ref.~\cite{bg}, is presented which provides a starting point to address the hierarchy problem and stability. Although our model does have a built-in stabilization mechanism due to a particular choice of UV boundary conditions,
a large hierarchy between the UV and IR scales can only be obtained by a significant amount of tuning. Nevertheless, different boundary conditions can be chosen which leave the IR scale undetermined, corresponding to a modulus field. Thus, new mechanisms may still be developed which solve the hierarchy problem in the soft-wall framework.

With these caveats aside we study bulk gauge fields and fermions in the soft-wall warped dimension. Even though the fifth dimension is infinite, the KK spectrum can be discrete, with a variety of spacing between resonances including linear Regge-like behavior.  The analysis of bulk fermions is particularly involved compared to the usual hard-wall setup. Specifically, it is necessary to go beyond the zero-mode approximation and fully account for the 5D Yukawa interactions that generate position-dependent fermion mass terms. 
The general problem with three generations requires a detailed numerical analysis, which we do not address in this paper. Instead, we illustrate in a simple single-generation model that many of the nice features of hard-wall models, such as fermion localization, mass hierarchies, and universal KK gauge couplings, occur in the soft-wall warped dimension. 

Finally, we construct electroweak models with custodial symmetry. We discuss the dynamics leading to an IR peaked bulk Higgs condensate responsible for breaking electroweak symmetry. Moreover, we find that electroweak constraints are not as stringent as their hard-wall counterparts, accommodating KK modes with masses of order the TeV scale.  
  
This paper is organized as follows: In Sec.~2, we discuss some of the general features of the soft-wall warped dimension. A dynamical 5D gravity model is then presented which provides a concrete realization of the soft-wall framework, and issues related to the separation of the UV and IR scales are discussed. We study bulk gauge fields and fermions in Sec.~3, fully accounting for the effect of the backreaction of the Higgs condensate on the fermion dynamics. In Sec.~4 we consider custodial electroweak models, studying the Higgs sector and electroweak constraints. Directions for future work and conclusions are presented in Sec. 5. 
   
\section{The soft-wall warped dimension}
\subsection{General features}
\label{sectiongeneral}

The basic feature which distinguishes the soft-wall warped dimension from the usual hard wall slice of
AdS is the replacement of the IR brane with a smooth spacetime cutoff. The metric describing 5D spacetime in the conformal coordinate $z$ can be written as
\begin{equation}
ds^2 = e^{-2 A(z)} \eta_{MN}dx^M dx^N~.
\label{swmetric}
\end{equation}
We will work with a pure AdS metric, $A(z)=\log{k z}$,  where $k$ is the AdS curvature scale and $\eta_{MN}={\rm diag}( - , + , + , + , +)$. In contrast to hard-wall models, the coordinate $z$ extends to infinity. The action describing the gauge and matter fields is 
\begin{equation}
S=\int d^5 x \sqrt{-g}\, e^{-\Phi} {\cal L}, 
\label{matteraction}
\end{equation}
where ${\cal L}$ is the matter field Lagrangian, and $\Phi$ is the ``dilaton''. Although we are taking a phenomenological approach, we have in mind that $\Phi$ is to be identified with the string theory dilaton and the action (\ref{matteraction}) may originate from some particular D-brane construction. 

The dilaton obtains a nontrivial background value $\Phi(z)$, providing a dynamical cutoff to spacetime and obviating the need for an IR brane. In the holographic picture, $\Phi$ is responsible for the confining dynamics at infrared energy scales. Indeed, we can identify an effective running coupling $g_5^2 e^{\Phi}\sim e^{\Phi}/N_c$, with $N_c$ the number of colors in the dual theory, which grows in the IR. Correspondingly, sources located at large $z$ will be strongly coupled, and processes involving exchange of IR localized bulk KK modes can become nonperturbative at high energies \cite{adsqcd, fpv}. However, for UV localized matter, as in the electroweak models that we will present, the effective description will remain perturbative sufficiently far into the infrared region.

Though there are many possible behaviors for the dilaton, we will only consider power-law behavior $\Phi(z) = (\mu z)^\nu$. In general the eigenfunctions of bulk fields with a power-law dilaton satisfy an analog 1D ``Schr\"{o}dinger'' equation with a power-law potential. A simple WKB approximation then shows that for large mode number $n$ the KK mass spectrum follows  
\begin{equation}
m_n^2 \sim \mu^2\, n^{2-2/\nu}. 
\end{equation}
Even though the conformal coordinate $z$ extends to infinity, for $\nu>1$ we obtain a discrete mass spectrum. 
In particular, for the case $\nu=2$ the spectrum exhibits linear ``Regge'' behavior. Later we will specialize to this case as it allows for analytic results. As $\nu\rightarrow \infty$ we recover the usual hard-wall mass spectrum $m_n^2\sim n^2$. The dilaton power-law
exponent, $\nu$, therefore provides a continuous parameter in which the KK mass spectrum varies
from a continuum to that associated with a compact extra dimension. As discussed in \cite{fpv}, there are other interesting but qualitatively distinct behaviors possible if $\nu \le 1$.  For example, a constant dilaton \cite{rs2} leads to ``unparticles'' \cite{un} from a 4D perspective, while ``hidden valley'' models \cite{hv} are obtained when $\nu=1$ \cite{adsun}. 

Though an IR brane is no longer needed, a UV boundary at small $z$ is still required in order to obtain the zero modes identified with the SM fields, which otherwise would not be normalizable. This also follows from holography, because typically the zero modes are (primarily) elementary fields associated with ``sources'' on the UV brane, rather than composites emerging from the dual gauge theory. As in hard-wall models, the UV brane will be located at a position $z_0= 1/k$.  

Note that there is an alternative way to model the soft wall, which relies on having an exponentially decaying metric \cite{sw2}. For many cases, the equations of motion are the same whether one uses a dilaton or the decaying metric, but differences can arise, in particular for bulk fermions and massive bosonic fields. We find in most cases that it is technically simpler to use a running dilaton as the soft wall with a pure AdS metric.

\subsection{A dynamical soft wall}
\label{sectiongrav}

Though it is possible to study certain aspects of soft-wall phenomenology from a purely bottom-up approach, a number of important questions cannot be addressed without reference to an underlying gravity theory. A dynamical gravitational model is required, for example, to address issues regarding generation of hierarchies and stability. 
In this section we present a dynamical 5D gravitational model which leads to a soft-wall warped dimension. 
The model is the same as that in Ref.~\cite{bg} with modifications to accommodate a UV boundary. 

Consider the Einstein frame action describing gravity and two scalar fields, the ``dilaton'' $\phi$ and the ``tachyon'' $T$:
\begin{eqnarray}
S & = & \int d^5x \sqrt{-g}~ \left( M^3 R - \frac{1}{2}g^{MN}  \partial_M \phi \partial_N \phi -  \frac{1}{2}g^{MN}\partial_M T \partial_N T -V(\phi,T) \right) \nonumber \\
&&- \int d^4x \sqrt{-g_{UV}} ~ \lambda_{UV}(\phi, T),
\label{aef} 
\end{eqnarray}
where $M$ is the 5D Planck scale. The bulk action contains a scalar potential $V(\phi, T)$, while the UV boundary located at $z_0=1/k$ is characterized by the induced metric $g_{UV}$ and boundary potential $\lambda_{UV}$. 

The solutions to (\ref{aef}) are most easily obtained through the introduction of a ``superpotential'' $W(\phi, T)$, which converts the system into a set of first-order differential equations~\cite{super0,superp}. Using this procedure, we can write the bulk and boundary potentials in the simple form
\begin{eqnarray}
V(\phi, T) & = & 18 \left[ \left( \frac{\partial W}{\partial \phi} \right)^2+\left( \frac{\partial W}{\partial T}\right)^2\right]-\frac{12}{M^3} W^2,   \label{pot1} \\
\nonumber \\
\lambda_{UV}(\phi, T) & = &6 \left[ W(\phi_{0}, T_{0})+\partial_\phi W(\phi_{0}, T_{0})(\phi-\phi_0)+ \partial_T  W(\phi_{0}, T_{0})(T-T_0)+\dots \right], \label{bpot0} \nonumber\\
\end{eqnarray}
where $\phi_0, T_0$ are the boundary values at $z=z_0$.
The extra terms in the boundary potential may contain higher powers of $(\phi-\phi_{0})$ and $(T-T_0)$ without affecting the background solution.

There exists a solution to the 5D gravity-dilaton-tachyon equations of motion with the metric $g_{MN}=e^{-2\widetilde{A}(z)}\eta_{MN}$ and the background solutions~\cite{bg}
\begin{eqnarray}
\widetilde{A}(z)&=&\frac{2}{3}(\mu z)^\nu+\log{kz}\,, \label{genA} \\
\phi(z)&=& \sqrt{\frac{8}{3}} M^{3/2}(\mu z)^\nu\,, \label{genphi} \\
T(z) & = &\pm 4 \sqrt{ 1+1/ \nu}\,M^{3/2} (\mu z)^{\nu/2}\,, \label{genT}
\end{eqnarray}
where the tilde in (\ref{genA}) distinguishes the Einstein frame from the ``string'' frame.
Note also that we have set the additive constants in the solutions (\ref{genphi}) and (\ref{genT}) to zero.
The superpotential which gives rise to this solution is
\begin{equation}
W(\phi, T)= M^3 k \left[(\nu+1) e^{T^2/(24(1+1/\nu)M^3)}  -\nu\left(1-\frac{\phi}{\sqrt{6}M^{3/2}}\right)e^{\phi/(\sqrt{6}M^{3/2})} \right],
\end{equation}
from which the scalar potential can be obtained using Eq.~(\ref{pot1}).

The parameter $\mu$ is an integration constant in the solution and sets the IR scale of the soft wall.
This is analogous to the radius in RS1, which is also identified as a modulus field. Without stabilization of the scale $\mu$, there should exist a massless radion associated with this modulus. However, we will see next that the UV boundary potential can in fact stabilize $\mu$, and we therefore expect that the radion becomes massive. A complete answer to this question can only be obtained by analyzing the fluctuations of the background solutions, which is beyond the scope of the present work.

The UV boundary conditions are found to be
\begin{eqnarray}
M^3\,e^{\widetilde{A}}\,\frac{\partial \widetilde{A}}{\partial z}\,\bigg\vert_{z_0}&=& W(\phi_0,T_0), \\
e^{\widetilde{A}}\,\frac{\partial \phi}{\partial z} \bigg\vert_{z_0}&=&6\,\partial_\phi W(\phi_0,T_0), \label{bcphi}\\
e^{\widetilde{A}}\,\frac{\partial T}{\partial z} \bigg\vert_{z_0}&=&6\,\partial_T W(\phi_0,T_0), \label{bcT}
\end{eqnarray}
which imply that 
\begin{eqnarray}
\phi_0 &=& \sqrt{\frac{8}{3}} M^{3/2}(\mu z_0)^\nu\,, \label{b0phi} \\
T_0 & = &\pm 4 \sqrt{ (1+1/ \nu)}\,M^{3/2} (\mu z_0)^{\nu/2}\, . \label{b0T}
\end{eqnarray}
Taking $z_0= 1/k$, Eqs. (\ref{b0phi}) and (\ref{b0T}) fix the soft-wall scale to be
\begin{equation}
\mu=k\left(\sqrt{\frac{3}{8}}\frac{\phi_0}{M^{3/2}}\right)^{1/\nu} =k \left(\frac{1}{\pm 4 \sqrt{1+1/\nu}}\frac{T_0}{M^{3/2}}\right)^{2/\nu}.
\label{mufix}
\end{equation}
Note that (\ref{mufix}) also implies a tuning between $\phi_0$ and $T_0$. Clearly, a large hierarchy cannot be generated between the UV scale $k$ and the soft-wall IR scale $\mu$ for $\nu>1$. Taking natural values for the boundary values, $\phi_0 \sim T_0 \sim M^{3/2}$ implies $\mu \lesssim k$, with a larger hierarchy for smaller values of $\nu$. In the case $\nu=2$ on which we will focus later, it is clearly not possible to generate the Planck-weak scale hierarchy without a significant amount of tuning. Interestingly, the hierarchy $\mu/k\sim 10^{-16}$ can be naturally generated for $\phi_0 \sim 0.1\, M^{3/2}$ and $\nu \sim 1/13$, but this does not give rise to a discrete KK particle spectrum. Nevertheless this deserves further study.
 
While the boundary action (\ref{bpot0}) fails to naturally generate a large hierarchy between $k$ and 
$\mu$, an alternative way to satisfy the boundary conditions for $\phi$ and $T$ is to let
\begin{equation}
\lambda_{UV}(\phi, T)  = 6 W(\phi, T). \label{bpot0s}  
\end{equation}
The boundary conditions following from the variational principle do not then fix the IR scale $\mu$. With this assumption other stabilization mechanisms can then be explored.
For example, we might consider an additional scalar field $S$, as in the Goldberger-Wise mechanism~\cite{gw}, with a small amplitude so that the backreaction on the metric can be neglected. If the field has a profile $S(z) \sim M^{3/2}(\mu z)^{\beta}$, and boundary condition analogous to those in (\ref{bcphi}) and (\ref{bcT}), this would suggest $\mu/k \sim \left(S_0 / M^{3/2} \right)^{1/\beta}$. A large hierarchy between $k$ and $\mu$ would be obtained if $0<\beta<1$. It would be interesting to look at the dynamics leading to this profile for $S$ and determine if the backreaction on the dilaton and metric can be made small.     

Although our main application of the soft-wall background will be to model electroweak physics, one can ask whether ordinary 4D gravity can be incorporated naturally into our model. The 4D Planck mass is given by
\begin{eqnarray} 
M_P^2&=&M^3 \int_{z_0}^\infty dz \,e^{-3\widetilde{A}(z)}, \nonumber \\
&=&\frac{2^{2/\nu}}{\nu} \frac{M^3 \mu^2}{k^3} \, \Gamma\left(-\frac{2}{\nu}\, , 2 \left(\frac{\mu}{k}\right)^\nu \right) 
\simeq \frac{M^3}{2k},
\label{mplanck}
\end{eqnarray}
where $\Gamma(n,x)$ is the incomplete Gamma function, and we have used $z_0 =1/k$ and assumed $\mu/ k \ll 1 $ in the last step. We can see that there is a problem because we would like to have $\mu\sim$ TeV to model electroweak physics. Lacking a robust mechanism that generates a hierarchy between $\mu$ and $k$ means that $k\sim\mu\sim$ TeV. If we take as usual $k \lesssim M$, then according to (\ref{mplanck}) we cannot account for the weakness of gravity.  

With these considerations, there are two possible cases for the UV scale $k\lesssim M$: (i) $k \ll M_P$, i.e. there is no large hierarchy and we project out the zero-mode graviton with Dirichlet conditions (for concreteness we will take $k\sim 1000 \mu$ as in \cite{little}); (ii) $k\sim M_P$, i.e.
we assume a suitable stabilization mechanism may be found and apply Neumann conditions to allow a massless graviton.  

Note that the metric (\ref{swmetric}) and action describing matter fields (\ref{matteraction}) is defined in the string frame, which is obtained by rescaling the dilaton $\phi=\sqrt{8/3}M^{3/2}\Phi$ and performing a conformal transformation $g_{MN}\rightarrow e^{-4\Phi/3}g_{MN}$. In the string frame, the background solutions for the metric and dilaton become
\begin{eqnarray}
A(z)&=&\log{kz},\, \label{genAs} \\
\Phi(z)&=& (\mu z)^\nu\,. \label{genphis} 
\end{eqnarray}
We have a pure AdS metric and power-law dilaton as advertised in Sec. \ref{sectiongeneral}.
Unless otherwise specified we will now restrict to $\nu=2$. This will give rise to a linear
Regge-like mass spectrum and will enable analytic solutions to be obtained. 
Other values of $\nu$ will lead to qualitatively similar  features.

\subsubsection{Graviton fluctuations}

Our explicit dynamical model allows us to study the properties of bulk graviton resonances, which may have interesting phenomenological implications if matter is located in the bulk. Consider the tensor fluctuations of the metric $g_{MN}$:
\begin{equation}
ds^2=e^{-2\widetilde{A}(z)}\big[ \left(\eta_{\mu\nu}+h_{\mu\nu}(x,z)\right)dx^\mu dx^\nu +dz^2 \big],
\end{equation}
where in the transverse-traceless gauge, $\partial_\mu h^{\mu\nu}= h^\mu_\mu = 0$, the 5D gravitational action becomes
\begin{equation}
S=M^3\int d^5x \sqrt{-g}R \rightarrow M^3\int d^5x \,e^{-3\widetilde{A}}\left(-\frac{1}{4}\partial_\rho h_{\mu\nu}\partial^\rho h^{\mu\nu} -\frac{1}{4}\partial_5 h_{\mu\nu}\partial_5 h^{\mu\nu}\right).
\end{equation}
The bulk graviton is expanded in KK modes
\begin{equation}
h_{\mu\nu}(x,z)=\sum_{n=0}^\infty h^n_{\mu\nu}(x)f_h^n(z),
\end{equation}
where the wavefunctions $f_h^n$ obey the equation of motion 
\begin{equation}
\partial_5 (e^{-3\widetilde{A}}\partial_5 f_h^n(z)) + m_n^2 e^{-3\widetilde{A}}f_h^n(z)=0,
\end{equation}
and the orthonormal condition
\begin{equation}
M^3 \int_{z_0}^\infty dz \,e^{-3\widetilde{A}(z)} f_h^n(z)f_h^m(z) = \delta^{nm}.
\label{normh}
\end{equation}
The normalization (\ref{normh}) leads to a canonical action for the graviton fluctuations. 
We must impose either Neumann or Dirichlet conditions on the wavefunctions at the UV boundary. 
Applying Neumann conditions $\partial_5 f_h^n(z_0)=0$ gives rise to a massless 4D graviton with wavefunction, $f^0_h=1/M_P$, that is UV localized with respect to a flat metric, where $M_P$ is defined in (\ref{mplanck}). Instead if Dirichlet conditions are applied then the zero mode is projected out. 

\begin{figure}
\centerline{\includegraphics[width=.8\textwidth]{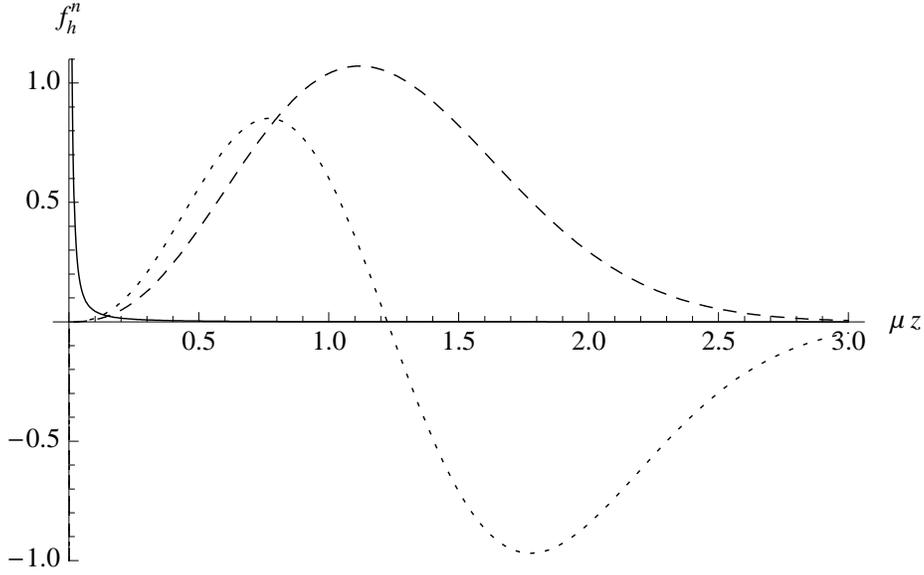}}
\caption{KK graviton profiles: The zero mode (solid), $n=1$ (dash), and $n=2$ (dot), for $\mu=1$ TeV and $k=1000$ TeV. If $k$ and $\mu$ have a Planck-weak scale separation, then the zero mode is further separated from the KK modes.}
\label{fig1}
\end{figure}

Next we consider the massive KK modes. It is convenient to make the redefinition $f_h^n(z)=e^{3\widetilde{A}(z)/2}\widehat{f}_h^n(z)$, which brings the equation of motion into the form of a 1D Schr\"{o}dinger equation:
\begin{equation}
\left[-\partial_5^2 + V(z)\right]\widehat{f}_h^n(z)=m_n^2 \widehat{f}_h^n(z),
\label{1dscheq}
\end{equation}
with the potential given by
\begin{equation}
V(z)=\frac{9}{4}\widetilde{A}'^2-\frac{3}{2}\widetilde{A}''=4\mu^4 z^2+4\mu^2+\frac{15}{4 z^2}.
\end{equation}
The normalizable solutions are given by
\begin{equation}
\widehat{f}_h^n(z)=N^n_h e^{-3 \widetilde{A}(z)/2} ~ U\left(-\frac{m_n^2}{8\mu^2}~,-1~,2 \mu^2 z^2 \right),
\end{equation}
where $U(a,b,y)$ is the Tricomi confluent hypergeometric function. The profiles $f_h^n(z)$
are therefore
\begin{equation}
 f_h^n(z)=N^n_h ~ U\left(-\frac{m_n^2}{8\mu^2}~,-1~,2 \mu^2 z^2 \right).
 \label{gravfn}
\end{equation}
The profiles with respect to a flat metric are plotted in Fig. \ref{fig1}.

The KK mass spectrum can be found by applying the UV boundary condition.
In the limit $\mu z_0 \ll 1$, the KK modes follow approximate linear trajectories:
\begin{equation}
m_n^2\simeq 8 \mu^2(n+1), \quad\quad n=1,2\dots.
\label{gravspect}
\end{equation}
The approximate mass formula (\ref{gravspect}) is valid for both Neumann and Dirichlet conditions. This is because the mass spectrum is largely determined by IR dynamics and is not overly sensitive to the UV boundary condition. 

For large $z$, the wavefunctions (\ref{gravfn}) are well approximated by Laguerre polynomials:
\begin{equation}
f^n_h(z)\simeq N^n_h (-1)^{n+1} (n-1)! ~  4 \mu^4 z^4 L^2_{n-1}\left( 2 \mu^2 z^2\right).
\label{gravwfn}
\end{equation}
Using (\ref{gravwfn}), we can derive an approximate expression for the normalization:
\begin{equation}
N^n_h \simeq \frac{(-1)^{n+1}}{(n+1)!}\frac{ k}{\mu}~\left[~\frac{ M^3 }{ k}
~\sum_{j,k=0}^{n-1}
~\frac{(-n+1)_j}{j! ~ \Gamma(j+3)}
~\frac{(-n+1)_k}{k! ~ \Gamma(k+3)} 
~\Gamma(j+k+3)\right]^{-1/2},
\end{equation}
where $\Gamma(x)$ is the gamma function and $(x)_n = \Gamma(x+n)/\Gamma(n)$ denotes the Pochhammer symbol. The sum can be performed,
\begin{equation}
\sum_{j,k=0}^{n-1}
~\frac{(-n+1)_j}{j! ~ \Gamma(j+3)}
~\frac{(-n+1)_k}{k! ~ \Gamma(k+3)} 
~\Gamma(j+k+3)=\frac{1}{n(n+1)},
\end{equation}
and using (\ref{mplanck}), we can write the normalization as
\begin{equation}
N^n_h\simeq\frac{(-1)^{n+1}}{M_P}\frac{k}{\mu}~\sqrt{\frac{2}{(n+1)!(n-1)!}}.
\end{equation}
Note that these results also follow from the analog 1D Schr\"{o}dinger equation (\ref{1dscheq}). In the limit
$z\rightarrow\infty$ the potential is equivalent to that of a harmonic oscillator with energy eigenvalues
(\ref{gravspect}) and eigenfunctions (\ref{gravwfn}).

As in hard-wall models, the couplings of the KK gravitons depend on where matter is located in the extra dimension. Later we will examine electroweak models with UV localized fermions. In this case the KK mode gravitons couple with a strength $f^n_h(z_0)\sim \mu/(k M_P)$, which is extremely suppressed and not likely to have observable consequences. This of course will change if fermions propagate in the bulk.

\section{Bulk fields}

We will now consider bulk gauge and fermion fields in the soft-wall background. As motivated in Sec. 2, the starting point will be the action (\ref{matteraction}) with an appropriate matter Lagrangian.

\subsection{Gauge Field}
Consider the simple case of a U(1) gauge field $A_M(x,z)$ in the bulk. The gauge field dynamics 
are described by the action
\begin{equation}
S = \int d^4x dz \sqrt{-g}\, e^{-\Phi}\left(-\frac{1}{4} F_{MN} F^{MN}\right).
\end{equation}
Performing a KK decomposition,
\begin{equation}
A_\mu(x,z)=\sum_{n=0}^\infty A_\mu^n(x)f_A^n(z),
\label{expa2}
\end{equation}
the wavefunctions obey the equation of motion
\begin{equation}
\partial_5 (e^{-(A+\Phi)}\partial_5 f_A^n)=-m_n^2e^{-(A+\Phi)}f_A^n,
\end{equation}
and are normalized according to 
\begin{equation}
\int_{z_0}^\infty dz \,e^{-(A+\Phi)} f_A^n(z) f_A^m(z) =\delta^{nm}.
\end{equation}
The profile of the massless mode is constant:
\begin{equation}
f_A^0(z)=\sqrt{-\frac{2k}{{\rm Ei}\left(-\mu^2 / k^2\right)}}\simeq \sqrt{\frac{k}{\log(k / \mu)-\gamma/2}}\,,
\label{0modeA}
\end{equation}
where ${\rm Ei}(x)$ is the exponential integral function, $\gamma\approx 0.577$ is the Euler-Mascheroni constant, and we have used $z_0=1/k$ and $\mu /k \ll 1$.
The wavefunctions of the massive modes are 
\begin{equation}
f_A^n(z)=N^n_A ~ U\left(-\frac{m_n^2}{4\mu^2}~,0~, \mu^2 z^2 \right).
\label{fan}
\end{equation}

Applying the Neumann condition to the wavefunctions at the UV boundary determines the mass spectrum of the excited vector modes. We find that in the limit $\mu/k \ll1$, the gauge boson masses follow approximate linear trajectories: 
\begin{equation}
m_n^2 \simeq 4\mu^2 n, \quad\quad n=1,2\dots.
\label{maapprox}
\end{equation}
For large $z$, the wavefunctions reduce to Laguerre polynomials:
\begin{equation}
f_A^n(z) \simeq N^n_A (-1)^{n+1} (n-1)! ~   \mu^2 z^2 L^1_{n-1}\left(  \mu^2 z^2\right).
\label{fappa}
\end{equation}
Similarly, as for the graviton wavefunction case, this form of the wavefunction can be used to derive an approximate expression for the normalization:
\begin{equation}
N^n_A \simeq \frac{(-1)^{n+1}}{n!}\sqrt{2 n k}\,.
\label{normaapp}
\end{equation}

\subsection{Fermions}
\label{fermions}

While the analysis of bulk gauge fields in the soft-wall background is straightforward, this is not the case
for fermions. Unlike in hard-wall models with an IR brane, the Higgs boson in a soft-wall background must necessarily propagate in the bulk. Since the Higgs profile should be peaked in the IR (to be dual to a composite electroweak symmetry breaking sector), the backreaction of the Higgs vacuum expectation value (VEV) on the bulk fermion equations of motion at large $z$ cannot be neglected. The correct way to proceed is to diagonalize the bulk equations of motion and obtain the SM fermion masses from the boundary conditions\footnote{Note that in general any model with a bulk Higgs condensate and bulk fermions should be analyzed in this way. However with a hard wall cutting off the extra dimension, it may be reasonable to treat the bulk Yukawa interaction as a perturbation and use the zero-mode approximation for fermions (although, see also \cite{cghnp}).}. This is different from the usual case in which fermions are analyzed using the zero-mode approximation, treating Yukawa interactions as perturbations and obtaining fermion masses from wavefunction overlap integrals~\cite{ferm,gp,huber}. Indeed, we show in Appendix~\ref{appferm} that the fermion zero-mode approximation in the soft-wall warped dimension has problems with strong coupling and normalizability. We therefore endeavor to fully account for the Higgs feedback on the fermion equations of motion. 

As we will see, the general analysis for three fermion generations is quite involved and requires a numerical approach that is beyond the scope of the present work to determine the mass eigenvalues and eigenvectors. Instead, we will specialize to a simple model with one generation and identical bulk masses for SU(2)$_L$ doublet and singlet SM fermions. This simple model allows for an analytical determination of masses and eigenfunctions and illustrates how in principle the general analysis can be done. More importantly, we show that some of the nice features of bulk fermions in hard-wall models, such as mass hierarchies and universal KK gauge couplings, which usually lead to the ``GIM'' mechanism, are also present in our single generation model. These features are likely to persist in a general setup with three fermion generations, allowing a complete treatment of flavor issues in the soft-wall background to be addressed.

Let us therefore begin with the general case of three fermion generations. Our conventions for fermions are summarized in Appendix \ref{conv}. In the bulk theory we have SU(2)$_L$ doublets $\Psi^{a i}_{L}(x,z)$ and singlets $\Psi^i_{R}(x,z)$, where $i$ is a flavor index and $a$ is a SU(2)$_L$ index. We define the two-component parts of the Dirac spinor as $\Psi^{a i}_{L\pm}=\pm \gamma^5 \Psi^{a i}_{L\pm}$ and similarly for $\Psi^i_R$. 
Neglecting for the moment Yukawa interactions, the dynamics of the bulk fermions is governed by the action
\begin{eqnarray}
S&=&-\int d^5x \sqrt{-g}~e^{-\Phi}\Big[~ \frac{1}{2}\left( \overline{\Psi}^{a i}_L e^M_A \gamma^A D_M \Psi^{a i}_L - D_M \overline{\Psi}^{a i}_L e^M_A \gamma^A \Psi^{a i}_L\right) + M_L^{ij}\overline{\Psi}^{a i}_L \Psi^{a j}_L  \nonumber \\
&&\qquad\qquad+\frac{1}{2}\left( \overline{\Psi}^i_R e^M_A \gamma^A D_M \Psi^i_R - D_M \overline{\Psi}^i_R e^M_A \gamma^A \Psi^i_R\right) + M_R^{ij}\overline{\Psi}^i_R \Psi^j_R \Big]~,
\label{ferma}
\end{eqnarray}
where $e^M_A=e^{A}\delta^M_A$ is the vielbein and $D_M=\partial_M +\omega_M$ is the covariant derivative with spin connection $\omega_M$. We work in a basis where the mass matrices $M_{L,R}$ are diagonal. 

In the absence of Yukawa interactions, we can obtain zero modes 
$\Psi^{ai(0)}_{L+}(x)$ and $\Psi^{i(0)}_{R-}(x)$ by applying Dirichlet conditions at the UV boundary to the fields $\Psi^{a i}_{L-}$ and $\Psi^i_{R+}$:
\begin{eqnarray}
\Psi^{a i}_{L-}(x,z) \bigg\vert_{z_0} &=&0,  \nonumber \\
\Psi^i_{R+}(x,z) \bigg\vert_{z_0} &=&0.
\label{bcdir}
\end{eqnarray}
However, with no IR boundary the Higgs boson must necessarily propagate in the bulk, significantly affecting the dynamics of the bulk fermions. Consider the bulk Yukawa interaction for the ``up-type'' fermions:
\begin{eqnarray}
S_{Yukawa}&=&-\int d^5 x \sqrt{-g} e^{-\Phi} \Big[\,\frac{\lambda^{ij}_5}{\sqrt{k}}\,\overline{\Psi}^{a i}_L(x,z) H^a(x,z)\Psi^j_R(x,z) +{\rm h.c.}\, \Big] \nonumber\\
&\equiv& -\int d^5 x \sqrt{-g} e^{-\Phi} \, \Big[ m^{ij}(z)\,\overline{\Psi}^i_L(x,z)\Psi^j_R(x,z) +{\rm h.c.}\, \Big],
\end{eqnarray}
where we have substituted the background value for the Higgs field 
$H(x,z)\rightarrow H(z)=\frac{h(z)}{\sqrt{2}} \footnotesize{\left(\begin{array}{c} 0\\1 \end{array}\right)}$ with the definition $\Psi_L\equiv\Psi_L^2$ and defined the effective mass term arising from the Yukawa interaction  
\begin{equation}
m^{ij}(z)\equiv \frac{\lambda^{ij}_5}{\sqrt{2\,k}}h(z).
\end{equation}

There is a $z$-dependent bulk mass mixing between $\Psi^i_L$ and $\Psi^j_R$ due to the Yukawa interaction. Defining $\Psi=e^{2A+\Phi/2}\psi$, the equations of motion for $\psi_{L\pm} $ and $\psi_{R\pm}$ are
\begin{eqnarray} 
\gamma^\mu \partial_\mu \psi^i_{L\pm} 
\mp \partial_5\psi^i_{L\mp}
 + e^{-A}M_L^{ij}\psi^j_{L\mp}
  + e^{-A}m^{ij}\psi^j_{R\mp} & = & 0, \\
  \gamma^\mu \partial_\mu \psi^i_{R\pm} 
\mp \partial_5\psi^i_{R\mp}
 + e^{-A}M_R^{ij}\psi^j_{R\mp}
  + e^{-A}m^{\dag ij} \psi^j_{L\mp} & = & 0. 
\end{eqnarray}
The KK expansions for $\psi^i_{L\pm}$ and $\psi^i_{R\pm}$ are
\begin{eqnarray}
\psi^i_{L\pm}(x,z)&=&\sum_{n=0}^\infty \psi^{i (n)}_{\pm}(x)f^{i (n)}_{L\pm}(z), \label{expf1} \\
\psi^i_{R\pm}(x,z)&=&\sum_{n=0}^\infty \psi^{i (n)}_{\pm}(x)f^{i (n)}_{R\pm}(z),
\label{expf2}
\end{eqnarray}
where $\gamma^\mu \partial_\mu \psi^{i(n)}_\pm =-m^i_n\psi^{i(n)}_\mp$. Defining the vectors
\begin{equation}
f^{i (n)}_{\pm}  =
\left( 
\begin{array}{c} 
 f^{i (n)}_{L\pm} \\
 f^{i (n)}_{R\pm}
\end{array} \right), 
\label{fvec} 
\end{equation}
the equations of motion for the wavefunctions $f_\pm^{i (n)}$ can be written in the compact form
\begin{equation}
\left[ \partial_5  \delta^{ij}   \pm  {\cal M}^{ij}\right] f^{j(n)}_{\pm}(z) =\pm m^i_n f^{i (n)}_\mp ,
\label{genferm}
\end{equation}
where the mixing matrix is defined as
\begin{equation}
{\cal M}=
e^{-A}\left( 
\begin{array}{cc} 
M^{ij}_L  & m^{ij}(z) \\
m^{\dag ij}(z)& M^{ij}_R 
\end{array} \right). 
\label{matrix} 
\end{equation}

The problem is now reduced to finding the profiles by solving Eq.~(\ref{genferm}), and determining the masses from the boundary conditions (\ref{bcdir}). This is a difficult problem owing to the coordinate dependence of the mass matrix, which implies that the transformation diagonalizing the mass matrix will in general be $z$-dependent. Such a $z$-dependent rotation will not leave the $\partial_5$-term invariant in (\ref{genferm}). This mixing is somewhat reminiscent of the ``twisted split fermion''
models of \cite{twisted}. One difference is that the Yukawa interactions mix SU(2)$_L$ doublet and singlet fermions, whereas the ``localizer'' scalar in split fermion models does not induce such a mixing. 

A new feature of this type of analysis is that we are really no longer searching for an exact zero-mode field but rather an ``almost zero mode'', which is to be identified with the SM fermion. The mass of the SM fermion is determined directly from the boundary conditions (\ref{bcdir}), rather than from an overlap integral.  To proceed further with a realistic three generation model of bulk fermions requires a numerical approach to solve the system (\ref{genferm}), which we will not pursue in this paper. Instead, we will now explore a simpler model with a single generation which can be solved analytically.

\subsubsection{One generation model}
\label{sectionmz}

With one generation of fermions ($i=1$), the mass mixing matrix (\ref{matrix}) becomes  
\begin{equation}
{\cal M}=
e^{-A}\left(
\begin{array}{cc} 
M_L  & m(z) \\
m(z)& M_R 
\end{array} \right), 
\label{matrix1} 
\end{equation}
where $m(z)$ can be taken to be real by a phase rotation of the fermions. In general, diagonalizing the matrix (\ref{matrix1}) still requires a $z$-dependent transformation. However, there is a special case when $M_L=M_R\equiv M$ that we now consider in which a global transformation diagonalizes the system. Defining  $f^{n}_\pm \equiv f^{1(n)}_\pm$ in (\ref{fvec}), we can diagonalize the equations of motion with the following transformation:
\begin{equation}
\left( 
\begin{array}{c} 
 g^{n}_{L\pm} \\
 g^{n}_{R\pm}
\end{array} \right)
=\frac{1}{\sqrt{2}}
\left( 
\begin{array}{cc} 
 1 & 1 \\
 1 & -1
\end{array} \right)
\left( 
\begin{array}{c} 
 f^{n}_{L\pm} \\
 f^{n}_{R\pm}
\end{array} \right).
\label{fvec1} 
\end{equation}

The equations of motion for the wavefunctions $g^{n}_{\{L,R\}\pm}$ are
\begin{eqnarray}
\left[\partial_5 \pm e^{-A} (M+m) \right]g^n_{L\pm}(z) &=& \pm m_n g^n_{L\mp}(z), \label{feom1}\\
\left[\partial_5 \pm e^{-A} (M-m) \right]g^n_{R\pm}(z) &=& \pm m_n g^n_{R\mp}(z), \label{feom2}
\end{eqnarray}
and are normalized according to 
\begin{equation}
\int_{z_0}^\infty dz ~\left(  g^n_{L\pm} g^m_{L\pm}+g^n_{R\pm}g^m_{R\pm} \right)=\delta^{nm}.
\label{normf2}
\end{equation}
From the UV boundary conditions (\ref{bcdir}) and the definitions (\ref{fvec1}), we can write the boundary conditions for the wavefunctions using the equations of motion (\ref{feom1}) and (\ref{feom2}):
\begin{eqnarray}
\label{bcprime0}
g^n_{L\pm} \bigg\vert_{z_0}& =& \pm g^n_{R\pm} \bigg\vert_{z_0}~,  \\
\big[\partial_5\pm e^{-A}(M+m)\big]g^n_{L\pm} \bigg\vert_{z_0}&=& \mp \big[\partial_5\pm e^{-A}(M-m)\big]g^n_{R\pm} \bigg\vert_{z_0}. 
\label{bcprime}
\end{eqnarray}

For a generic Higgs background $h(z)$ it will not be possible to obtain analytic forms for the wavefunctions. Let us therefore specialize to a concrete example of a Higgs profile in which the wavefunctions can be found analytically. We will assume the Higgs profile is given by 
\begin{equation}
h(z)=\eta k^{3/2} \mu^2 z^2,   
\label{higgsvev}
\end{equation}
where $\eta$ is a dimensionless ${\cal O}(1)$ coefficient\footnote{Note that an analytic solution with a linear VEV has been considered in Ref.~\cite{adsun}, but assuming a KK expansion (\ref{expf1}) and (\ref{expf2}) does not give rise to a discrete spectrum.}. Later in Sec.~4.1 we will analyze the dynamics leading to this Higgs profile. The first-order equations (\ref{feom1}) and (\ref{feom2}) can each be decoupled, allowing us to write the following second-order differential equations:
\begin{eqnarray}
\left[-\partial_5^2 +V_{\{L,R\}\pm}(z) \right]g^n_{\{L,R\}\pm}(z) & = & m_n^2 g^n_{\{L,R\}\pm}(z).
\label{feommassive}
\end{eqnarray}
The ``Schr\"{o}dinger'' potentials are given by
\begin{eqnarray}
V_{L\pm}(z)&=&b^2 \mu^4 z^2+(2c \mp 1)b \, \mu^2+\frac{c(c\pm 1)}{z^2}, \nonumber \\
V_{R\pm}(z)&=&b^2 \mu^4 z^2-(2c \mp 1)b \, \mu^2+\frac{c(c\pm 1)}{z^2},
\label{schrod}
\end{eqnarray}
where we have used $M=ck$ and defined the parameter $b=\lambda_5 \eta/\sqrt{2}$ in the potentials (\ref{schrod}). 
The solutions to (\ref{feommassive}) that are finite at large $z$ are
\begin{eqnarray}
g^n_{L+}(z)&=&N^n_{L+}\, e^{ - \, b \,\mu^2\, z^2/2}z^{-c} ~U \left( -\frac{m_n^2}{4 b \mu^2}~ ,~ \frac{1}{2}- c ~ , \,  \, b \, \mu^2 z^2 \right), \\
g^n_{L-}(z) &=& N^n_{L-}\, e^{ - \, b \,\mu^2\, z^2/2}z^{1-c} ~U \left( -\frac{m_n^2}{4 b \mu^2}+1~ ,~ \frac{3}{2}- c ~ , \,  \, b \, \mu^2 z^2 \right), \\
g^n_{R+}(z)&=&N^n_{R+}\, e^{ - \, b \,\mu^2\, z^2/2}z^{1+c} ~U \left( -\frac{m_n^2}{4 b \mu^2}+1~ ,~ \frac{3}{2}+ c ~ , \,  \, b \, \mu^2 z^2 \right), \\
g^n_{R-}(z) &=& N^n_{R-}\, e^{ - \, b \,\mu^2\, z^2/2}z^{c} ~U \left( -\frac{m_n^2}{4 b \mu^2}~ ,~ \frac{1}{2}+ c ~ , \,  \, b \, \mu^2 z^2 \right).
\end{eqnarray}

The fermion mass spectrum is obtained by applying the boundary conditions (\ref{bcprime0}) and (\ref{bcprime}), which yields the following equation:
\begin{eqnarray}
\frac{1}{4}m^2_n z_0^2\,
U \left( -\frac{m_n^2}{4 b \mu^2}+1~ ,~ \frac{3}{2}- c ~ , \,  \, b \, \mu^2 z_0^2 \right)
U \left( -\frac{m_n^2}{4 b \mu^2}+1~ ,~ \frac{3}{2}+ c ~ , \,  \, b \, \mu^2 z_0^2 \right) \nonumber \\
-\, U \left( -\frac{m_n^2}{4 b \mu^2}~ ,~ \frac{1}{2}- c ~ , \,  \, b \, \mu^2 z_0^2 \right)
U \left( -\frac{m_n^2}{4 b \mu^2}~ ,~ \frac{1}{2}+ c ~ , \,  \, b \, \mu^2 z_0^2 \right)=0.
\label{fspectrum}
\end{eqnarray}
The first massive mode is to be identified with the SM fermion, so it is of interest to determine its mass. In the limit $\mu z_0 \ll1$, and assuming the first mode is light $m_0^2/(4b\mu^2) \ll 1$, an expansion of Eq. (\ref{fspectrum}) reveals a very light mode for $|c|>1/2$:
\begin{equation}
m_0^2 \simeq \frac{2 b \mu^2}{\Gamma(-1/2+|c|)}(b \mu^2 z_0^2)^{-1/2+|c|}.  
\end{equation}
In the regime $-1/2< c <1/2$, we find instead that the fermion mass is of order $b\mu^2$:
\begin{equation}
m_0^2\simeq\frac{4 b \mu^2}{\pi \sec c\pi -\psi(1/2-c)-\psi(1/2+c)}~,
\end{equation}
where $\psi$ is the digamma function.
Thus we see that it is possible to generate a small fermion mass (e.g. electron) or a large mass (e.g. top quark) by choosing different values of the bulk mass parameter $c$, at least in this simple case of one generation. 

\begin{figure}
\centerline{\includegraphics[width=0.7\textwidth]{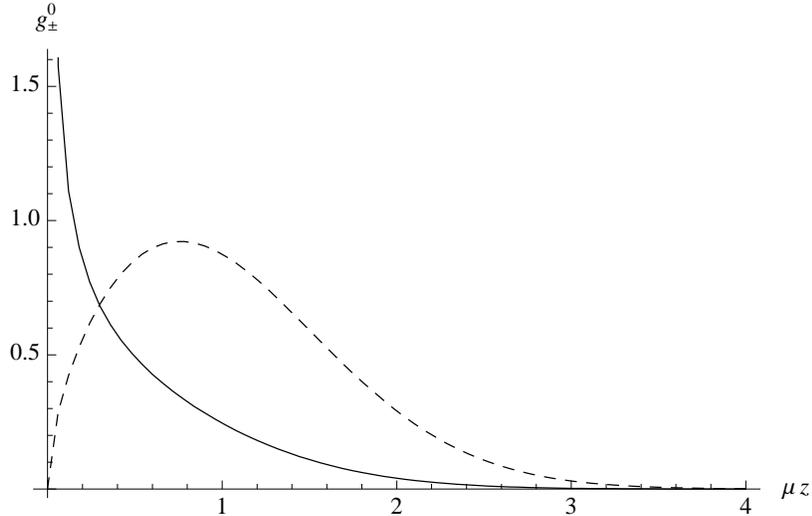}}
\caption{ The fermion profiles of the lightest mass eigenstates $\psi^0_\pm$ for $b=1, c=1/2, \mu=1$ TeV and $k=1000$ TeV. The solid (dashed) line indicates $g^0_{+}(z)$ ($g^0_{-}(z)$).}
\label{fig3}
\end{figure}

Note that given our assumption that the bulk Dirac fermions have the same mass $M=c k$, one fermion is always UV localized while the other is IR localized. This can be seen by examining the wavefunctions of the lightest mode $\psi^0_\pm$, which are obtained from the normalization condition (\ref{normf2}) and defined as
\begin{equation}
g^0_{\pm}(z)=\sqrt{(g^0_{L\pm}(z))^2 +(g^0_{R\pm}(z))^2}. 
\label{fermg0}
\end{equation}
We have plotted these profiles in Fig.~\ref{fig3} for $c=1/2$, in which case the left-handed mode is UV localized while the right-handed mode is peaked out into the fifth dimension. 

\begin{figure}
\centerline{\includegraphics[width=0.9\textwidth]{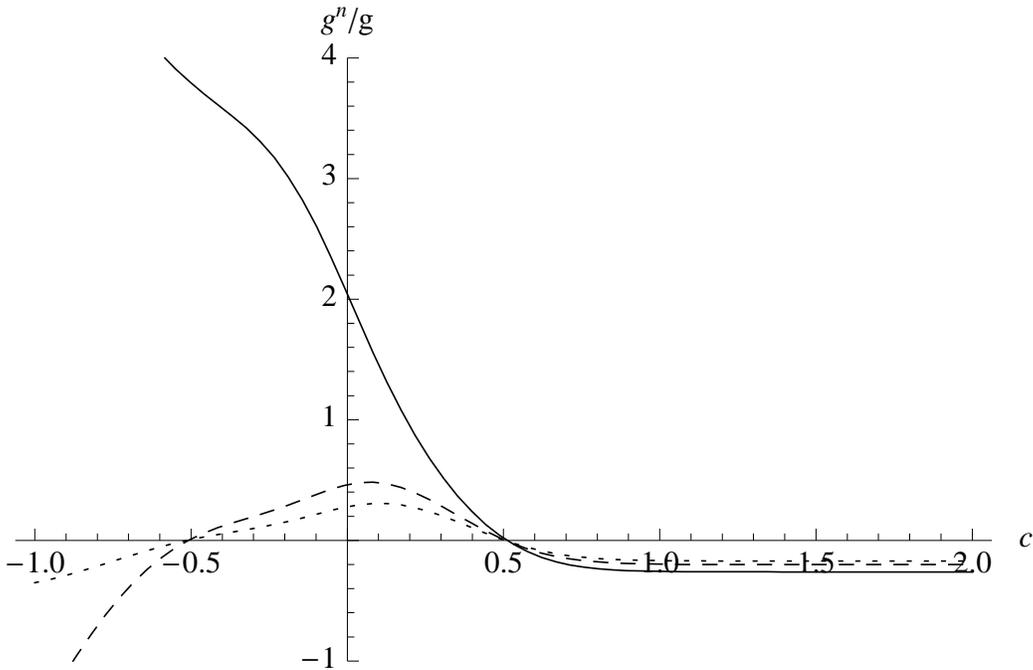}}
\caption{The ratio $g^{n}/g$ for the $n=1$ (solid), $n=2$ (dashed), and $n=3$ (dotted) KK gauge modes with $b=1, \mu=1$ TeV, and $k=1000$ TeV.}
\label{fig4}
\end{figure}

We have seen that it is possible to generate a mass hierarchy, at least in this simple bulk fermion model. It is interesting to ask whether any of the other nice features of flavor physics present in the hard-wall models also appear in this simple soft-wall setup. For example, the usual hard-wall framework with bulk fermions contains a built in ``GIM'' mechanism suppressing FCNC induced by the exchange of KK gauge bosons. 
This is due to the fact that light fermions are UV localized and to a good approximation couple universally to these excited KK gauge modes~\cite{gp}.
Let us investigate the coupling of, say, $\psi^0_+(x)$ to these excited states, to obtain the dependence on the localization parameter $c$ of the zero mode. For simplicity, we will consider fermions with the same charge coupled to a U(1) gauge boson (i.e. photon). The bulk gauge coupling is given by
\begin{equation}
S= i g_5 \int d^5x \sqrt{-g}\,e^{-\Phi}\,\left[\,
\overline{\Psi}_L e^M_A \gamma^A A_M \Psi_L 
+\left( L \rightarrow R \right) \right] .
\end{equation}
Inserting the KK decompositions (\ref{expa2}), (\ref{expf1}), and (\ref{expf2}) and using (\ref{fvec1}), the 4D gauge coupling for $\psi^0_+$ is found to be
\begin{equation}
g=g_5 f^0_A \simeq g_5 \sqrt{\frac{k}{\log(k/\mu)-\gamma/2}}.
\end{equation}
Similarly, the coupling of two zero-mode fermions to a KK gauge boson is given by
\begin{equation}
g^{n}=g_5\int_{z_0}^\infty dz~ g^0_+(z)f^n_A(z)g^0_+(z),
\label{gn}
\end{equation}
where the wavefunctions are obtained from (\ref{fan}) and (\ref{fermg0}).
In Fig.~\ref{fig4} we plot the ratio $g^{n}/g$ as a function of $c$. We can see that, as in the hard-wall models with bulk fermions, the couplings quickly become universal for UV localized fermions, $c>1/2$, while for $c<1/2$ the couplings are larger and non-universal. 

Finally, we point out that in this simple model in which the different fermions have the same bulk mass, only one of the fermions $\psi^0_+$ or $\psi^0_-$ may enjoy a universal coupling due to their opposite localization (see Fig. \ref{fig3}).  Clearly, the relevant case to consider is when fermions have different bulk masses.
The analytic results we have obtained in this single-generation model are promising, and suggest that the flavor puzzle in the SM may be addressed in the soft-wall framework. In this light, the general problem discussed above clearly deserves further consideration.

\section{Electroweak models}
\label{sectionew}

In this section we investigate electroweak models with custodial symmetry \cite{custodial} in the soft-wall background. For simplicity, we will consider SM fermions localized on the UV brane, which are interpreted as elementary states in the holographic theory. It should be possible to generalize these models to include bulk fermions based on the analysis presented in Sec.~\ref{fermions}. We will focus on bulk gauge fields interacting with a Higgs field peaked at large $z$, which is dual to electroweak symmetry breaking via strong dynamics.   
 
Consider a bulk gauge theory with an SU(2)$_L \times$SU(2)$_R \times $U(1)$_X$ custodial 
symmetry. On the UV boundary the gauge symmetry is broken via boundary conditions to the electroweak subgroup SU(2)$_L \times$U(1)$_Y$. In the IR region, the custodial symmetry is broken to the vector subgroup via a bulk Higgs condensate. In the usual hard-wall setups, this symmetry breaking is achieved via a Higgs localized on the IR brane \cite{custodial} or via IR-brane boundary conditions as in Higgsless models \cite{higgsless}. In fact, our setup is very similar to the ``gaugephobic Higgs'' model~\cite{gphobic}, though with a different background geometry and no IR brane.

The Lagrangian of the model is given by
\begin{eqnarray}
S&=& \int d^5 x \sqrt{-g}\,e^{-\Phi}\Bigg[-\frac{1}{4g_5^2}L_{MN}^a L^{aMN}-\frac{1}{4g_5^2}R_{MN}^a R^{aMN}-\frac{1}{4g_5^{'2}}X_{MN} X^{MN}  \nonumber \\
&&\qquad\qquad -\,{\rm Tr}|D_M H|^2 -V(H)\Bigg] - \int d^4 x \sqrt{-g_{UV}} e^{-\Phi}  \, V_{UV}(H) ,
\label{ewl}
\end{eqnarray}
where $L^a_M(x,z)$, $R^a_M(x,z)$, and $X_M(x,z)$ represent SU(2)$_L$, SU(2)$_R$, and U(1)$_X$ gauge fields, respectively. In addition there is a bulk Higgs boson $H(x,z)$ with bulk and boundary potentials $V(H)$ and $V_{UV}(H)$, respectively. We have chosen the right- and left-handed gauge couplings to be equal for simplicity. 

The gauge fields satisfy the following UV boundary conditions,
\begin{equation}
\begin{array}{rcr}
\partial_5 L^a_\mu\bigg\vert_{z_0} = 0,  && 
R^{1,2}_{\mu}\bigg\vert_{z_0} =  0,  \\
\partial_5 \left( \frac{1}{g_5^{'2}} X_\mu +\frac{1}{g_5^2} R_\mu^3 \right)\bigg\vert_{z_0}= 0,& & 
(X_\mu- R^3_\mu)\bigg\vert_{z_0}  = 0,
\label{UVbc}
\end{array}
\end{equation}
which break SU(2)$_R \times$U(1)$_X \rightarrow$U(1)$_Y$.
The bulk Higgs fields is a bidoublet under SU(2)$_L \times$SU(2)$_R$, and acquires a nontrivial profile along the fifth direction:
\begin{equation}
\langle H(z) \rangle=\frac{h(z)}{\sqrt{2}} 
\left( \begin{array}{cc} 
  1 & 0 \\
  0 & 1  
\end{array} \right). 
\label{higgspro}
\end{equation}
This profile breaks SU(2)$_L\times$SU(2)$_R \rightarrow$ SU(2)$_V$. We therefore define vector and axial-vector fields $V,A=(L\pm R)/\sqrt{2}$, with wavefunctions $v(p,z)$ and $a(p,z)$ satisfying the equations of motion:
\begin{eqnarray}
\partial_5 \left(e^{-(A+\Phi)}\partial_5  v(p,z)\right) & = & p^2 e^{-(A+\Phi)} v(p,z), \label{veceom} \\
\partial_5 \left(e^{-(A+\Phi)}\partial_5 a(p,z)\right) -e^{-(3A+\Phi)}g_5^2 h^2(z) a(p,z) &=&p^2 e^{-(A+\Phi)}a(p,z). 
\label{axeom} 
\end{eqnarray}
The vector profile $v(p,z)$ is obtained from (\ref{fan}), while the exact form of the axial-vector profile can only be determined after specifying the Higgs VEV $h(z)$. We will next consider two simple cases which allow for an analytical determination of $a(p,z)$. Note that the $X$ gauge boson profile is also given by $v(p,z)$. 

From a 4D perspective, the theory contains a massless photon, a KK tower of charged $W$ bosons, and a KK tower of neutral $Z$ bosons with the lightest states in these towers identified with the SM $W$ and $Z$ bosons, respectively. To determine the mass spectra, we apply the UV boundary conditions in Eq. (\ref{UVbc}). For the $W$ tower, the spectrum (with $m_n^2=-p^2$) is determined by the following equation:
\begin{equation}
v(p,z_0)a'(p,z_0)+a(p,z_0)v'(p,z_0)=0, \label{wmass}
\end{equation}
while for the neutral $Z$ boson tower we find
\begin{equation}
v'(p,z_0)\left[ g_5^2 (v(p,z_0)a'(p,z_0)+a(p,z_0)v'(p,z_0))+2 g_5'^2  v(p,z_0)a'(p,z_0)\right]=0. \label{zmass}
\end{equation}
The prime ($'$) in Eqs. (\ref{wmass}) and (\ref{zmass}) denotes differentiation with respect to $z$. Note that of the two equations in (\ref{zmass}), one equation ($v'(p,z_0)=0$) corresponds to the excited modes of the photon, while the other equation determines the KK spectrum of the $Z$ boson.

To match the 5D theory to the 4D effective theory, we can relate the parameters $g_5$, $g_5'$, and $\mu$ to, for instance, the electric charge and the masses of the $W$ and $Z$ bosons determined from (\ref{wmass}) and (\ref{zmass}). The electric charge is computed from the normalization of the photon wavefunction, contained in the $L^3$, $R^3$, and $X$ bulk gauge bosons, and is given by
\begin{equation}
e^2\simeq \frac{g_5^2 g_5'^2}{g_5^2+2 g_5'^2} \,\frac{k}{\log(k/\mu)-\gamma/2} ~.
\end{equation}
The $W$ and $Z$ boson masses will be computed for specific Higgs profiles below, but first we consider the dynamics of the Higgs sector and present a simple model leading to an IR-peaked Higgs background profile.

\subsection{Higgs dynamics}

We now analyze the dynamics leading to a bulk Higgs condensate. An understanding of the Higgs dynamics is important for more than just aesthetic reasons; any realistic phenomenological study requires a concrete dynamical model to analyze the Higgs fluctuations and determine, for example, the mass of the physical Higgs scalar and its couplings to SM fields.

The bulk Higgs potential in (\ref{ewl}) is assumed to have the form
\begin{equation}
V(H)=m_H^2(z) {\rm Tr}|H|^2, 
\end{equation}
where we have defined a $z$-dependent effective mass
\begin{equation}
m_H^2(z)=k^2\left[\alpha(\alpha-4)-2\alpha\mu^2 z^2\right].
\label{higgsmz}
\end{equation}
According to the AdS/CFT dictionary, the particular constant mass-squared in (\ref{higgsmz}) corresponds to an operator with dimension $\Delta_H= |\alpha-2|+2$ in the dual theory. The $z$-dependent mass term is assumed to arise from a coupling to another scalar field which obtains a background VEV. In fact, in our gravity model there are two candidates for these scalar fields, the dilaton $\Phi$ 
and tachyon $T$. Interaction terms like $\Phi |H|^2$ or $T^2 |H|^2$ can provide the $z^2$ part of the mass term, although we do not need to specify the precise origin of this term for the phenomenological analysis. Note also that there is a tuning between the different terms in (\ref{higgsmz}).

Inserting the background (\ref{higgspro}), we find the following equation of motion for $h(z)$:
\begin{equation}
\partial_5 (e^{-(3A+\Phi)}\partial_5 h) - e^{-(5A+\Phi)}m_H^2(z) h=0.
\label{higgseom1}
\end{equation}
The general solution to this equation is 
\begin{equation}
h(z)=z^\alpha\left(c_0 +  c_1 \Gamma\left(2-\alpha, -\mu^2 z^2\right) \right),
\end{equation}
where $c_0,c_1$ are arbitrary constants. 
Demanding finiteness of this solution in the soft-wall background implies $c_1=0$, which leads to
\begin{equation}
h(z)=c_0 z^\alpha. 
\label{hC}
\end{equation}

We must add a UV boundary potential to ensure that the solution (\ref{hC}) can nontrivially satisfy the boundary condition. An appropriate choice is 
\begin{equation}
V_{UV}(H) =\frac{\lambda_0}{k^2} \left( {\rm Tr} |H|^2 -v_0^2 \right)^2,
\end{equation}
which leads to the UV boundary condition
\begin{equation}
\left(\partial_5 h -\frac{2\lambda_0}{k^2} h (h^2-v_0^2) \right)\Bigg\vert_{z_0}=0.
\end{equation}
Substituting (\ref{hC}) into this boundary condition gives rise to two possible solutions, a trivial solution $c_0=0$, as well as a nontrivial solution: 
\begin{equation}
c_0^2=k^{3+2\alpha}\left( \frac{v_0^2}{k^3}+\frac{\alpha}{2 \lambda_0} \right).
\label{c}
\end{equation}
It is not sufficient for a nontrivial solution to merely exist; we must also determine if it is the vacuum state. This can be done by computing the energy density per unit brane volume, $\cal H $, analogous to the calculation performed in Ref.~\cite{higgsH}. Consequently, the difference between the trivial and nontrivial solution energy densities is found to be
\begin{equation}
{\cal H}(h=0)-{\cal H}(h=c_0 z^\alpha ) = \frac{1}{\lambda_0}k^4 e^{-\mu^2 z_0^2}\left( \frac{\lambda_0 v_0^2}{k^3}+\frac{\alpha}{2} \right)^2.
\end{equation}
Therefore, the nontrivial Higgs background (\ref{c}) will be the ground state provided that this difference is positive, which occurs when $\lambda_0>0$. Incidentally the energy density of the nontrivial solution is of order
$k^4 + v_0^2 k$, so provided $v_0^2 \lesssim k^3$ and $k \lesssim M$ the backreaction on the gravitational background can be neglected.

In order to accomplish electroweak symmetry breaking, the Higgs profile should ``turn on'' in the IR near $z\sim 1/\mu$, suggesting that $c_0 \propto k^{3/2}\mu^\alpha$. We thus require that
\begin{equation}
\frac{v_0^2}{k^3}+\frac{\alpha}{2 \lambda_0} \sim \left( \frac{\mu}{k} \right)^{2\alpha}.
\label{uvtune}
\end{equation}
This is clearly tuned, since the quantity on the left hand side is naturally of order one. The need for this tuning is due to the fact that the stabilizing potential is located on the UV brane. Eq.~(\ref{uvtune}) suggests two possible situations: either $v^2_0$ is small and $\lambda_0$ is large, or a partial cancellation occurs between the two terms on the left-hand side of (\ref{uvtune}), in which case both $v^2_0$ and $\lambda_0$ can have perturbative values. To determine which case can be realized we 
need to consider the fluctuations of the Higgs background.

To analyze the Higgs fluctuations let $h(z)\rightarrow h(z)+\widetilde{h}(x,z)$. The equation of motion for $\widetilde{h}$ is
\begin{equation}
 e^{-(3A+\Phi)}\Box \widetilde{h}+ \partial_5 (e^{-(3A+\Phi)}\partial_5\widetilde{h}) - e^{-(5A+\Phi)}m_H^2 \widetilde{h}=0.
\end{equation}
Due to the boundary quartic potential, the UV boundary condition for the fluctuation is a nonlinear equation for which an analytic solution is difficult to obtain. Instead an approximate solution can be found by performing a linearized fluctuation analysis. In this case the boundary condition for the fluctuation becomes
\begin{equation}
\left(\partial_5 -\frac{2\lambda_0}{k^2}\left( (h^2-v_0^2) +2h^2 \right)\right)\widetilde{h} \Bigg\vert_{z_0}=0.
\end{equation}
Expanding the fluctuation as 
\begin{equation}
 \widetilde{h}(x,z)=\sum_{n=1}^\infty \widetilde{h}^n(x)g^n_{\widetilde{h}}(z),
\end{equation}
and defining $g^n_{\widetilde{h}}(z)=e^{(3A+\Phi)/2}\,\widehat{g}^n_{\widetilde{h}}(z)$, the profiles $\widehat{g}^n_{\widetilde{h}}(z)$ satisfy a Schr\"{o}dinger equation with the potential
\begin{equation}
V_{\widetilde{h}}(z)=\mu^4 z^2 +2(1-\alpha) \mu^2 + ((\alpha-2)^2-1/4)\frac{1}{z^2}.
\end{equation}
The solutions for the (unhatted) profiles are then
\begin{equation}
 g_{\widetilde{h}}^n(z)=  N_{\widetilde{h}}^n z^\alpha U\left(-\frac{m_n^2}{4\mu^2}~,\alpha-1~, \mu^2 z^2 \right),
\end{equation}
where $ N_{\widetilde{h}}^n$ is a normalization factor.
Applying the boundary conditions, the Higgs mass spectrum is determined by the equation
\begin{equation}
m_n^2 z_0^2 \,U\left(-\frac{m_n^2}{4\mu^2}+1~,\alpha~, \mu^2 z_0^2 \right)
-4\,\zeta \,U\left(-\frac{m_n^2}{4\mu^2}~,\alpha-1~, \mu^2 z_0^2 \right)=0,
\end{equation}
where $\zeta=\alpha+2\lambda_0v_0^2/k^3\sim 2\lambda_0 (\mu/k)^{2\alpha}$.
In the limit $|\zeta|\ll 1$ the Higgs (lowest lying mode) mass-squared is $m_0^2\simeq 2\zeta k^2/\log(k/\mu)$. 
For $\zeta<0$ we find a tachyon mode, and a zero mode at $\zeta=0$, so we restrict to $\zeta>0$. 
The Higgs mass increases as we increase $\zeta$. Note that these results are at the linearized level
and the nonlinear terms in the UV boundary condition have been neglected.

Earlier we argued that $\lambda_0>0$ if the nontrivial Higgs profile is to be the vacuum state of the theory. 
Now we see that this condition also implies that there are no tachyon modes provided $v_0^2/k^3 >-\alpha/(2\lambda_0)$. In particular, for $v_0^2/k^3 = -\alpha/(2\lambda_0) +\epsilon$ then (\ref{uvtune}) can be satisfied with $\epsilon \sim (\mu/k)^{2\alpha}$, implying that $v_0^2$ and $\lambda_0$ can have perturbative values.
Thus, a perturbative solution describing electroweak symmetry breaking with a light Higgs boson can be found. However, for large enough $\zeta$, corresponding to a heavy Higgs or technicolor limit, the theory becomes nonperturbative. 

\subsection{Electroweak constraints}
With fermions localized on the UV brane and a bulk custodial symmetry, the most important constraint on this model comes from the $S$ parameter \cite{pt}. Of course, one would like to extend fermions into the bulk in a realistic manner to understand the SM flavor structure. In this case, there are other constraints that arise from loop level contributions to the $T$ parameter from KK mode fermions and nonuniversal corrections to the $Z\overline{b}b$ coupling \cite{custodial,ewrs}, as well as stringent constraints from flavor violation \cite{flavor}. Mechanisms to weaken these constraints have been developed recently, (e.g. using different custodial representations for third generation fermions \cite{zzb}), and there is no reason to expect such mechanisms cannot be implemented in the soft-wall warped framework. Nevertheless, the constraint from $S$ is still fairly restrictive in hard-wall models, forcing the KK scale to be around 3 TeV \cite{custodial}. It is thus interesting to see whether or not the constraint from $S$ can be weakened in a soft-wall background.

Recently, Ref.~\cite{fpv} found that the KK scale can indeed be lowered depending on the assumptions regarding the type of soft wall and Higgs condensate. In particular, they considered an example with a ``linear'' soft wall ($\nu=2$ in our notation) with a quadratic Higgs profile, finding that the KK scale can be around 2 TeV. We will verify this result, and present another example for the linear soft wall in which the constraints are even less severe. 
 
To calculate the $S$ parameter we will use the boundary effective action approach \cite{S1} which is particularly convenient when fermions are UV localized. Following \cite{S1, S2}, the general expression for the vector and axial-vector self energies is
\begin{eqnarray}
\Sigma_V &= &-\frac{1}{g_5^2} e^{-(A+\Phi)}\frac{\partial_5 v}{v}\bigg\vert_{z_0}, \\
\Sigma_A& = &-\frac{1}{g_5^2} e^{-(A+\Phi)}\frac{\partial_5 a}{a}\bigg\vert_{z_0}. 
\end{eqnarray}
The $S$ parameter is defined as
\begin{equation}
S=8 \pi(\Sigma_V'(0)-\Sigma_A'(0))~.
\end{equation}
From the exact expression for the vector profile given in (\ref{fan}), the vector self energy is
\begin{equation}
\Sigma_V(p^2)=\frac{e^{-\mu^2 z_0^2}}{2 g_5^2 k} p^2 \frac{ U \left( 1+\frac{p^2}{4\mu^2} \, , 1\, , \mu^2 z_0^2 \right)}{U\left(\frac{p^2}{4\mu^2}~,0~, \mu^2 z_0^2 \right)}.
\end{equation}
In the limit $\mu z_0\ll 1$ we find 
\begin{equation}
\Sigma'_V(0) \approx \frac{1}{2 g_5^2 k}\left( -\gamma-2 \log{\mu z_0}\right).
\label{vecder}
\end{equation}

We now examine two explicit examples of profiles $h(z)$ which allow for an analytic determination of the axial-vector profile $a(p,z)$ and $S$, finding in each case that the KK scale can be lowered.

\subsubsection{Linear VEV}
Assuming $\nu=2$ the first case we consider is when the Higgs VEV is linear in $z$, so that
\begin{equation}
g_5^2 h^2(z)=\xi k^2 \mu^2 z^2,
\label{linearvev}
\end{equation}
where $\xi$ is a dimensionless parameter. This requires choosing $\alpha=1$ or $m_H^2(z)=
-3 k^2-2\mu^2 z^2$. In the dual holographic theory this corresponds to electroweak symmetry breaking 
with an operator of dimension $\Delta_H=3$.
From the equation of motion (\ref{axeom}), we find the axial-vector profile $a(z)$:
\begin{equation}
a(p,z)=U\left(\frac{p^2}{4\mu^2}+\frac{\xi}{4}~,0~, \mu^2 z^2 \right).
\end{equation}
By expanding the spectrum equations (\ref{wmass}) and (\ref{zmass}) in the limit $\mu z_0 \ll 1$, $\xi \ll1$ we find two light modes that can be identified with the $W$ and $Z$ bosons, with masses:
\begin{eqnarray}
m^2_W&\approx & \frac{1}{2} \xi \mu^2,  \\
m^2_Z&\approx & \frac{1}{2} \frac{g_5^2+2g_5'^2}{g_5^2+g_5'^2} \xi \mu^2.
\end{eqnarray}
We can see the custodial symmetry at work in the relationship between the $W$ and $Z$ masses~\cite{higgsless}. The ratio $m_W^2/m_Z^2 \approx (g_5^2+g_5'^2)/(g_5^2+2g_5'^2)\simeq g^2/(g^2+g'^2)$, where $g,g'$ are the 
SU(2)$_L$, U(1)$_Y$ gauge couplings, respectively.

The closed form expression for the axial-vector self energy is
\begin{equation}
\Sigma_A(p^2) =\frac{e^{-\mu^2 z_0^2}}{2 g_5^2 k} ( p^2+\xi \mu^2 ) \frac{ U \left( 1+\frac{p^2}{4\mu^2}+\frac{\xi}{4} \, , 1\, , \mu^2 z_0^2 \right)}{U\left(\frac{p^2}{4\mu^2}+\frac{\xi}{4}~,0~, \mu^2 z_0^2 \right)}.
\end{equation}
Taking the limit $\mu z_0\ll 1$, $\xi \ll 1$ the derivative becomes 
\begin{equation}
\Sigma'_A(0) \approx \frac{1}{2 g_5^2 k}\left( -\gamma-2 \log{\mu z_0}-\frac{\pi^2}{12} \xi \right). \label{axder}
\end{equation}
Combining (\ref{axder}) with (\ref{vecder}), we find the $S$ parameter for the case of a linear Higgs VEV:
\begin{equation}
S \approx \frac{\pi ^3 \xi}{3 g_5^2 k}
\simeq \frac{2\pi^3}{3 g^2( \log(k/\mu)-\gamma/2)} \frac{m_W^2}{\mu^2}. 
\label{Slinear}
\end{equation}
Requiring $S < 0.2$ implies that when the UV scale is 1000 TeV, the IR scale $\mu \leq 0.5\, {\rm TeV}$.
Thus, the first KK gauge boson resonances have masses of order 1 TeV. If we choose the UV scale to be of order $M_P$ then the constraint is even weaker, and the first KK modes can be quite light, of order 300-500 GeV! Also, since the spacing between successive modes is 2$\mu$, in this scenario it may actually be possible to observe the linear trajectory at the LHC, although this requires a further detailed phenomenological study.

\subsubsection{Quadratic VEV}
Next for $\nu=2$ we consider a quadratic profile for the Higgs,
\begin{equation}
g_5^2 h^2(z)=\xi k^2 \mu^4 z^4.
\label{quadraticvev}
\end{equation}
This requires choosing $\alpha=2$ or $m_H^2(z)=-4 k^2-4\mu^2 z^2$. In the dual holographic theory this corresponds to electroweak symmetry breaking with an operator of dimension $\Delta_H=2$.
The axial-vector profile is then
\begin{equation}
a(p,z)=e^{\mu^2 z^2 (1-\sqrt{1+\xi})/2}\,U\left(\frac{p^2}{4\mu^2\sqrt{1+\xi}}~,0~, \sqrt{1+\xi}\,\mu^2 z^2 \right).
\end{equation}
Expanding (\ref{wmass}) and (\ref{zmass}) in the limit $\mu z_0 \ll 1$, $\xi \ll 1$, the masses of the $W$ and $Z$ bosons are found to be
\begin{eqnarray}
m^2_W&\approx& \frac{1}{4}\frac{1}{\log(k/\mu) -\gamma/2} \xi \mu^2,  \\
m^2_Z&\approx& \frac{1}{4} \frac{g_5^2+2g_5'^2}{g_5^2+g_5'^2} \frac{1}{\log(k/\mu) -\gamma/2} \xi \mu^2.
\end{eqnarray}
The axial-vector self energy can then be computed and is given by,
\begin{equation}
\Sigma_A(p^2) =\frac{e^{-\mu^2 z_0^2}}{2 g_5^2 k} \left[ p^2 \frac{ U \left( 1+\frac{p^2}{4\mu^2\sqrt{1+\xi}} \, , 1\, , \sqrt{1+\xi} \mu^2 z_0^2 \right)}{U\left(\frac{p^2}{4\mu^2 \sqrt{1+\xi}}~,0~, \sqrt{1+\xi}\mu^2 z_0^2 \right)} -2\mu^2 (1-\sqrt{1+\xi})\right], 
\end{equation}
which leads to the expression for $\Sigma'(0)$ in the limit $\mu z_0 \ll1$:
\begin{equation}
\Sigma'_A(0) \approx \frac{1}{2 g_5^2 k}\left( -\gamma-2 \log{\mu z_0}-\log{\sqrt{1+\xi}} \right). \label{axder2}
\end{equation}
The $S$ parameter is therefore given by 
\begin{equation}
S\approx \frac{2 \pi}{g_5^2 k}\log{(1+\xi)} \simeq \frac{8 \pi}{ g^2} \frac{m_W^2}{\mu^2}.
\end{equation}
In the case when the Higgs VEV is quadratic in $z$, the constraint $S<0.2$ translates into an upper bound of $\mu \leq 1.3$ TeV, which is very similar to the result obtained in Ref.~\cite{fpv} for the same mass term (using their $\epsilon =1$). There is some weak dependence on the ratio $k/\mu$ in $S$, and taking $k \sim M_p$, the lower bound on $\mu$ becomes approximately $ 1.2 $ TeV.

\section{Discussion and conclusions}
The soft-wall warped dimension generalizes the usual hard-wall framework used to model electroweak
physics. A power-law dilaton is responsible for providing a smooth spacetime 
cutoff and corresponds to breaking conformal symmetry with an operator of finite dimension in the 
holographic dual theory. While the dilaton plays a similar role to that encountered in string theory and 
D-brane configurations, we presented a bottom-up dynamical solution with gravity and two scalar fields. 
This solution provides a backdrop for electroweak model building and lays the groundwork for investigations into important issues related to the gravitational sector.

For instance, it is still an open question whether the hierarchy problem can be solved in the soft-wall warped dimension with $\nu>1$. Within our 5D gravity model, a particular set of UV boundary conditions fixes the IR scale $\mu$, but we were able to show that a large hierarchy between the curvature scale $k$ and the IR scale $\mu$ could not be obtained naturally. 
An alternative choice in UV boundary conditions leaves the scale $\mu$ undetermined, corresponding to a modulus field, and 
therefore allows for the possibility of turning on an additional Goldberger-Wise type scalar field~\cite{gw}. It would be interesting to perform a detailed scalar fluctuation analysis of our solution and check whether introducing a Goldberger-Wise field could lead to a naturally large hierarchy between $k$ and $\mu$.

Leaving aside the gravitational issues we then considered the properties of bulk fields in the soft-wall background. Both the zero-mode graviton and gauge fields have constant profiles and therefore become UV localized with respect to a flat metric. Even though the warped dimension is infinite we showed that there exists a normalizable and discrete KK spectrum. This is qualitatively distinct from the original Randall-Sundrum model with the IR brane removed. Moreover the KK spacing between resonances depends on the dilaton power-law exponent and allows for a variety of possible behavior. In particular, for bulk gravitons and gauge fields a linear Regge-like spectrum (as in QCD) can be obtained. 

On the other hand the phenomenology of bulk fermions was not so straightforward. 
In the soft-wall framework a bulk Higgs condensate leads to $z$-dependent fermion masses,
which grow in the IR. Thus, the bulk Yukawa interaction cannot be treated as a perturbation and the full backreaction must be taken into account. The general case for three fermion generations will likely require a numerical analysis. However, we were able to analytically solve a special single-generation case, obtaining hierarchies in fermion masses as well as universal KK mode gauge couplings depending on fermion localization.
These results warrant the continued investigation of a complete soft-wall analog of the usual hard-wall flavor models.

Our simple soft-wall background setup also allowed the Higgs dynamics to be analyzed. We were able to show that the specific bulk Higgs condensate leading to fermion localization is a vacuum state of the theory and a linearized fluctuation analysis confirmed that there are no tachyonic modes. 
Interestingly the form of the bulk Higgs mass required to obtained the specific $z$-dependent fermion mass term follows from a coupling $T^2|H|^2$, suggesting that the ``tachyon" scalar field $T$ plays a crucial role in generating fermion masses. 
The specific form of the bulk Higgs condensate was then used to analyze an electroweak model with custodial symmetry and UV-localized fermions. 
The $S$ parameter was computed analytically, and it was shown that for various background Higgs VEVs, electroweak constraints are not as stringent compared to hard-wall models, with KK masses of order the TeV scale. 

The analysis presented in this paper can be used to study other qualitatively different possibilities. 
Most of the results obtained concerning bulk fields and electroweak physics focused on the case where the dilaton exponent $\nu=2$, corresponding to linear trajectories for the KK states. This case is analytically tractable, but we see no issue with numerically analyzing other values of $\nu$.  For example, when $\nu<1$ the ratio $\mu/k$ can be made naturally small and this would be an interesting case to explore.

As mentioned in Sec. 2, there is a strong coupling issue because the effective 5D coupling grows in the IR. The effective description therefore remains valid up to some large $z$ cutoff which can be made arbitrarily large for $N_c\rightarrow \infty$ in the dual theory. In fact, the extra dimension could be truncated at some large $z$
i.e. have both a hard and soft wall. 
As long as the hard wall is located sufficiently far into the extra dimension
many of the phenomenological features of the soft wall will be preserved. For example, the lowest lying resonances can still have exotic power-law mass trajectories that would eventually transition to the usual hard-wall spectrum. It would therefore be worth studying such a setup with both a running dilaton and a IR brane.   

Finally, the detailed collider phenomenology of the soft-wall Standard Model relevant for the LHC remains to be done.
In particular, with the KK scale being somewhat lower than in hard-wall models as well as the couplings between SM fermions and the resonances being somewhat weaker, the phenomenology could be qualitatively distinct from that in hard-wall models. Furthermore it would be worthwhile to generalize the electroweak models constructed in Sec.~4 by placing fermions in the bulk, accounting for the nonconstant bulk mass terms generated from Yukawa interactions. A fully realistic model of flavor, incorporating all electroweak constraints, will provide an interesting alternative to the usual hard-wall setups and deserves further study.

\section*{Acknowledgements}
We thank Alex Pomarol for useful discussions. This work was supported in part by a Department of Energy grant DE-FG02-94ER40823 at the University of Minnesota, and an award from Research Corporation. T.G. is also supported by the Australian Research Council.


\appendix
\def\theequation{\thesection.\arabic{equation}}
\setcounter{equation}{0}
\section{Fermion conventions}
\label{conv}

In this Appendix we present our conventions for fermions \cite{Weinberg}. The 5D Clifford algebra is 
\begin{equation}
\{ \gamma_M,\gamma_N \}=2~\eta_{MN}=2~{\rm diag}(-,+,+,+,+).
\end{equation}
We take as a basis the following gamma matrices:
\begin{eqnarray}
\gamma^\mu = -i \left(
\begin{array}{cc}
0 & \sigma^\mu \\
\overline{\sigma}^\mu & 0
\end{array}\right),
& &
\gamma^5 =  \left(
\begin{array}{cc}
1 & 0 \\
0 & -1
\end{array}
\right),
\end{eqnarray}
where $\sigma^\mu = (1, \sigma^i)$ and $\sigma^i$ are the usual Pauli matrices. Note that with this basis the proper Dirac conjugate is defined as $\overline{\Psi}=\Psi^\dagger i\gamma^0$. 

To deal with fermions in curved spacetime, we must introduce the vielbein $e^A_M$, defined through the relation
\begin{equation}
g_{MN}=e^A_M e^B_N \eta_{AB}.
\end{equation}
The covariant derivative is defined as $D_M = \partial_M +\omega_M$, where $\omega_M$ is the spin connection:
\begin{equation}
\omega_M = \frac{i}{2}{\cal J}_{AB}~\omega^{AB}_M.
\end{equation}
The Lorentz generators ${\cal J}_{AB}$ are given by
\begin{equation}
{\cal J}_{AB}=-\frac{i}{4}~\left[\gamma^A,\gamma^B\right],
\end{equation} 
and so the spin connection can be written as
\begin{equation}
\omega_M =\frac{1}{8}\omega_{M A B}~ \left[\gamma^A,\gamma^B\right].
\label{scon}
\end{equation} 
The coefficients ${{\omega_M}^A}_B$ are determined by
\begin{equation}
{{\omega_M}^A}_B= e^A_R ~ e^S_B ~\Gamma^R_{MS} - e^R_B ~\partial_M e^A_S,
\end{equation}
where $\Gamma^R_{MS}$ is the Christoffel symbol.

Specializing to the case of a conformal metric $g_{MN}=e^{-2A(z)}\eta_{MN}$, the vielbein is given by
\begin{equation}
e^A_M=e^{-A(z)}\delta^A_M,
\end{equation}
and the spin connection is found to be
\begin{equation}
\omega_M = \left(-\frac{A'}{2}\gamma_\mu \gamma^5, 0 \right).
\end{equation}

\section{Zero-mode approximation}
\label{appferm}

In this Appendix we will discuss problems with the zero-mode approximation for bulk fermions in the soft-wall background. We focus on fermions with a constant bulk mass $M=c k$, neglecting Yukawa interactions, and discuss the potential issues with strong coupling and zero-mode normalizability. These issues depend on how one chooses to model the soft wall, either with the $z^2$ asymptotics in the dilaton or instead in the metric. 

The problems discussed below are ultimately related to the fact that the fifth dimension extends to $z\rightarrow \infty$. If we consider Yukawa interactions with a bulk Higgs, the IR peaked Higgs profile can considerably alter the dynamics of bulk fermions, avoiding the problems discussed in this Appendix.

\subsection{Dilaton soft wall}

First, we can imagine the dilaton providing the soft wall, with $\Phi(z)=\mu^2 z^2$ and a pure AdS metric $A(z)=\log{k z}$. The fermion profiles obey an equation of motion analogous to (\ref{feom1}) with $m(z)=0$:
\begin{equation}
\big[\partial_5 \pm e^{-A} M \big]g^n_\pm(z) = \pm m_n g^n_\mp(z).
\label{eomfa1}
\end{equation}
The zero modes have a power-law profile:
\begin{equation}
g^{0}_{\pm}(z)\propto z^{\mp c}.
\end{equation}
These modes are normalizable if $1\mp2c <0$, meaning only UV localized zero modes are allowed.

Let us examine the gauge coupling between two (+) zero-mode fermions and a KK gauge boson, given in Eq. (\ref{gn}):
\begin{eqnarray}
g^{n}&\propto& g_5 \int^\infty_{z_0} dz \, z^{-2c}~U\left(-\frac{m_n^2}{4\mu^2}~,0~, \mu^2 z^2 \right),\nonumber \\
&\simeq & g_5 \int_{z_0}^\infty dz \, z^{-2c}z^{m_n^2 /2\mu^2},
\label{g0n0a}
\end{eqnarray}
where we have used the asymptotic large $z$ behavior of the hypergeometric function,
\begin{equation}
U(a,b,y)\sim y^{-a},
\end{equation}
in the final step. Noting the mass spectrum (\ref{maapprox}), this coupling becomes
\begin{equation}
g^{n} \propto \int_{z_0}^\infty dz\, z^{2n-2c},
\end{equation}
which diverges for $n>c-1/2$. Therefore, once a particular $c$ value  is chosen, the coupling $g^{n}$ diverges for sufficiently large gauge boson KK mode number $n$. 

\subsection{Metric soft wall}

It is also possible to model the soft wall with an exponentially decaying metric, with $\widetilde{A}(z)=2\mu^2 z^2/3 +\log{k z}$. 
Again, we can follow the analysis in Sec.~\ref{sectionmz}, this time setting $\Phi=m(z)=0$, and replacing $A(z)\rightarrow \widetilde{A}(z)$. We then obtain the equation of motion for the fermion profiles:
\begin{equation}
\big[\partial_5 \pm e^{-\widetilde{A}} M \big]g^n_\pm(z) = \pm m_n g^n_\mp(z).
\label{eomfa}
\end{equation}
The massless mode solutions can be obtained straightforwardly by integrating Eq. (\ref{eomfa}), leading to
\begin{equation}
g^0_{\pm}(z)\propto e^{\mp c ~ {\rm Ei}\left(-2\mu^2 z^2/3\right)/2}.
\end{equation}
However, this solution is not normalizable. The exponential integral function, vanishes as $z\rightarrow \infty$, and thus the profile $g^0_\pm(z)$ approaches a constant at large $z$. Noting the normalization condition (\ref{normf2}), we see that the zero mode is not normalizable, and is therefore absent from the theory.


\newpage

\begin{thebibliography}{99}

\bibitem{rs}
  L.~Randall and R.~Sundrum,
  Phys.\ Rev.\ Lett.\  {\bf 83}, 3370 (1999)
  [arXiv:hep-ph/9905221].

\bibitem{gb1}
  H.~Davoudiasl, J.~L.~Hewett and T.~G.~Rizzo,
  Phys.\ Lett.\  B {\bf 473}, 43 (2000)
  [arXiv:hep-ph/9911262].

\bibitem{gb2}
  A.~Pomarol,
  Phys.\ Lett.\  B {\bf 486}, 153 (2000)
  [arXiv:hep-ph/9911294].

\bibitem{ferm}
  Y.~Grossman and M.~Neubert,
  Phys.\ Lett.\  B {\bf 474}, 361 (2000)
  [arXiv:hep-ph/9912408].

\bibitem{chang}
  S.~Chang, J.~Hisano, H.~Nakano, N.~Okada and M.~Yamaguchi,
  Phys.\ Rev.\  D {\bf 62}, 084025 (2000)
  [arXiv:hep-ph/9912498].

\bibitem{gp}
  T.~Gherghetta and A.~Pomarol,
  Nucl.\ Phys.\  B {\bf 586}, 141 (2000)
  [arXiv:hep-ph/0003129].
 
\bibitem{huber}
  S.~J.~Huber and Q.~Shafi,
  Phys.\ Lett.\  B {\bf 498}, 256 (2001)
  [arXiv:hep-ph/0010195].

\bibitem{gim}
  K.~Agashe, G.~Perez and A.~Soni,
  Phys.\ Rev.\  D {\bf 71}, 016002 (2005)
  [arXiv:hep-ph/0408134].

\bibitem{adscft}
  J.~M.~Maldacena,
  Adv.\ Theor.\ Math.\ Phys.\  {\bf 2}, 231 (1998)
  [Int.\ J.\ Theor.\ Phys.\  {\bf 38}, 1113 (1999)]
  [arXiv:hep-th/9711200];
  S.~S.~Gubser, I.~R.~Klebanov and A.~M.~Polyakov,
  Phys.\ Lett.\  B {\bf 428}, 105 (1998)
  [arXiv:hep-th/9802109];
  E.~Witten,
  Adv.\ Theor.\ Math.\ Phys.\  {\bf 2}, 253 (1998)
  [arXiv:hep-th/9802150].

\bibitem{holo}
  N.~Arkani-Hamed, M.~Porrati and L.~Randall,
  JHEP {\bf 0108}, 017 (2001)
  [arXiv:hep-th/0012148];
  R.~Rattazzi and A.~Zaffaroni,
  JHEP {\bf 0104}, 021 (2001)
  [arXiv:hep-th/0012248];
  M.~Perez-Victoria,
  JHEP {\bf 0105}, 064 (2001)
  [arXiv:hep-th/0105048].

\bibitem{review}
  T.~Gherghetta,
  arXiv:hep-ph/0601213.
  
\bibitem{adsqcd}
  A.~Karch, E.~Katz, D.~T.~Son and M.~A.~Stephanov,
  Phys.\ Rev.\  D {\bf 74}, 015005 (2006)
  [arXiv:hep-ph/0602229].
   
\bibitem{fpv}
  A.~Falkowski and M.~Perez-Victoria,
  arXiv:0806.1737 [hep-ph].
  
\bibitem{bg}
  B.~Batell and T.~Gherghetta,
  Phys.\ Rev.\  D {\bf 78}, 026002 (2008)
  [arXiv:0801.4383 [hep-ph]].

\bibitem{rs2}
  L.~Randall and R.~Sundrum,
  Phys.\ Rev.\ Lett.\  {\bf 83}, 4690 (1999)
  [arXiv:hep-th/9906064].
  
\bibitem{un}
  H.~Georgi,
  Phys.\ Rev.\ Lett.\  {\bf 98}, 221601 (2007)
  [arXiv:hep-ph/0703260].

\bibitem{hv}
  M.~J.~Strassler and K.~M.~Zurek,
  Phys.\ Lett.\  B {\bf 651}, 374 (2007)
  [arXiv:hep-ph/0604261].
  

\bibitem{adsun}
  G.~Cacciapaglia, G.~Marandella and J.~Terning,
  arXiv:0804.0424 [hep-ph].

\bibitem{sw2}
  O.~Andreev,
  Phys.\ Rev.\  D {\bf 73}, 107901 (2006)
  [arXiv:hep-th/0603170].

\bibitem{super0}
  K.~Skenderis and P.~K.~Townsend,
  Phys.\ Lett.\  B {\bf 468}, 46 (1999)
  [arXiv:hep-th/9909070].

\bibitem{superp}
  O.~DeWolfe, D.~Z.~Freedman, S.~S.~Gubser and A.~Karch,
  Phys.\ Rev.\  D {\bf 62}, 046008 (2000)
  [arXiv:hep-th/9909134].

  \bibitem{gw}
  W.~D.~Goldberger and M.~B.~Wise,
  Phys.\ Rev.\ Lett.\  {\bf 83}, 4922 (1999)
  [arXiv:hep-ph/9907447].
  
\bibitem{little}
  H.~Davoudiasl, G.~Perez and A.~Soni,
  Phys.\ Lett.\  B {\bf 665}, 67 (2008)
  [arXiv:0802.0203 [hep-ph]].
  
\bibitem{cghnp}
 S.~Casagrande, F.~Goertz, U.~Haisch, M.~Neubert and T.~Pfoh,
  arXiv:0807.4937 [hep-ph].

\bibitem{twisted}
Y.~Grossman, R.~Harnik, G.~Perez, M.~D.~Schwartz and Z.~Surujon,
Phys.\ Rev.\ D {\bf 71}, 056007 (2005) 
[arXiv:hep-ph/0407260]. 

\bibitem{custodial}
  K.~Agashe, A.~Delgado, M.~J.~May and R.~Sundrum,
  JHEP {\bf 0308}, 050 (2003)
  [arXiv:hep-ph/0308036].

\bibitem{higgsless}
  C.~Csaki, C.~Grojean, H.~Murayama, L.~Pilo and J.~Terning,
  Phys.\ Rev.\  D {\bf 69}, 055006 (2004)
  [arXiv:hep-ph/0305237];
  C.~Csaki, C.~Grojean, L.~Pilo and J.~Terning,
  Phys.\ Rev.\ Lett.\  {\bf 92}, 101802 (2004)
  [arXiv:hep-ph/0308038].
  
\bibitem{gphobic}
  G.~Cacciapaglia, C.~Csaki, G.~Marandella and J.~Terning,
  JHEP {\bf 0702}, 036 (2007)
  [arXiv:hep-ph/0611358].

\bibitem{higgsH}
  F.~Coradeschi, S.~De Curtis, D.~Dominici and J.~R.~Pelaez,
  JHEP {\bf 0804}, 048 (2008)
  [arXiv:0712.0537 [hep-th]];
  C.~P.~Burgess, C.~de Rham and L.~van Nierop,
  arXiv:0802.4221 [hep-ph].
   
\bibitem{pt}
  M.~E.~Peskin and T.~Takeuchi,
  Phys.\ Rev.\ Lett.\  {\bf 65}, 964 (1990);
  M.~E.~Peskin and T.~Takeuchi,
  Phys.\ Rev.\  D {\bf 46}, 381 (1992).


\bibitem{ewrs}
  M.~S.~Carena, E.~Ponton, J.~Santiago and C.~E.~M.~Wagner,
  Nucl.\ Phys.\  B {\bf 759}, 202 (2006)
  [arXiv:hep-ph/0607106].


\bibitem{flavor}
  M.~Bona {\it et al.}  [UTfit Collaboration],
  JHEP {\bf 0803}, 049 (2008)
  [arXiv:0707.0636 [hep-ph]];
  K.~Agashe {\it et al.},
  Phys.\ Rev.\  D {\bf 76}, 115015 (2007)
  [arXiv:0709.0007 [hep-ph]];
  C.~Csaki, A.~Falkowski and A.~Weiler,
  arXiv:0804.1954 [hep-ph].

\bibitem{zzb}
  K.~Agashe, R.~Contino, L.~Da Rold and A.~Pomarol,
  Phys.\ Lett.\  B {\bf 641}, 62 (2006)
  [arXiv:hep-ph/0605341].

\bibitem{S1}
  R.~Barbieri, A.~Pomarol and R.~Rattazzi,
  Phys.\ Lett.\  B {\bf 591}, 141 (2004)
  [arXiv:hep-ph/0310285].
  
\bibitem{S2}
  K.~Agashe, C.~Csaki, C.~Grojean and M.~Reece,
  JHEP {\bf 0712}, 003 (2007)
  [arXiv:0704.1821 [hep-ph]].

\bibitem{Weinberg}
S.~Weinberg,
``The Quantum theory of fields. Vol. 1: Foundations,''
{\it  Cambridge, UK: Univ. Pr. (1995) 609 p};
``The quantum theory of fields.  Vol. 3: Supersymmetry,''
{\it  Cambridge, UK: Univ. Pr. (2000) 419 p}.  

\end{thebibliography}
\end{document}